\documentclass{amsart}

\usepackage{epsfig}
\usepackage{graphicx}
\usepackage{amsmath,amssymb,amsfonts,latexsym}
\usepackage{graphicx}

\usepackage{amssymb}
\usepackage{enumerate, xspace}
\usepackage{dsfont}
\usepackage{afterpage}
\usepackage{changebar}
\usepackage{amsthm}
\usepackage{rotating}

\numberwithin{equation}{section}
\begin{document}

\title[van der Waals interaction in polymerase-beta DNA replication]
{Nucleotide insertion initiated by van der Waals interaction during polymerase beta \\DNA replication}
\author[Andrew Das Arulsamy]{Andrew Das Arulsamy} %\email{sadwerdna@gmail.com}

%\address{Institute of Mathematical Sciences, University of Malaya, 50603 Kuala Lumpur, Malaysia}
%\address{Present Address: Condensed Matter Group, Division of Interdisciplinary Science, F-02-08 Ketumbar Hill, Jalan Ketumbar, 56100 Kuala-Lumpur, Malaysia}


\date{\today}

\begin{abstract}
We give unambiguous theoretical analyses and show that the exclusive biochemical reaction involved in a single nucleotide insertion into the DNA primer can be efficiently tracked using the renormalized van der Waals (vdW) interaction of a stronger type, the Hermansson blue-shifting hydrogen bond effect, and the Arunan composite hydrogen-vdW bond. We find that there are two biochemical steps involved to complete the insertion of a single base (cytosine) into the 3$'$ end of a DNA primer. First, the O3$'$ (from a DNA primer) initiates the nucleophilic attack on P$_{\alpha}$ (from an incoming dCTP), in response, O3$_{\alpha}$ (bonded to P$_{\alpha}$) interacts with H$'$ (bonded to O3$'$). These interactions are shown to be strongly interdependent and require the forming and breaking of P---O and H---O covalent bonds, which in turn imply that we do not need any external energy supply. 
\end{abstract}

\maketitle
~~\\ \\
\textbf{Affiliation}: Condensed Matter Group, Division of Interdisciplinary Science, F-02-08 Ketumbar Hill, Jalan Ketumbar, 56100 Kuala-Lumpur, Malaysia\\ \\
\textbf{E-mail}: sadwerdna@gmail.com\\ \\
\textbf{Tel.}: +603-91333546\\ \\
\textbf{Fax}: Not available \newpage

\section{Introduction}

The reaction pathways involving polymerase(pol)-$\beta$ in eukaryotic cells, which have high fidelity and specificity are presently one of the most subtle scientific information needed to elucidate what constitute the origin of pol-$\beta$ specificity and fidelity~\cite{rad,alb}. We chose the eukaryotic enzyme pol-$\beta$ here for two obvious reasons--- it is one of the simplest and smallest (39 kDa) known DNA polymerase enzyme, and it has been well-studied both experimentally and \textit{via} simulations since the 1970s~\cite{ul}. Radhakrishnan and Schlick~\cite{rad}, and Alberts, Wang and Schlick~\cite{alb} have carried out the constrained quantum/molecular mechanical calculations to obtain the relevant pol-$\beta$ DNA replication pathways with the lowest energies. Here, ``constrained'' means certain variables are computationally made not-to-float, following the crystal structure data reported by Batra \textit{et al}.~\cite{bat}, and from their earlier studies~\cite{alb}. The constrained variables include the position and the numbers of water molecules and ions (Mg$^{2+}$, Na$^+$ and Cl$^-$), the charge states and the conformational changes of the pol-$\beta$ residues (D190 (aspartate 190) and others) and the incoming dCTP (deoxyribocytosine 5$'$-triphosphate), and so on~\cite{alb}. The crucial result~\cite{rad,alb} is that the initial deprotonation of the terminal DNA primer $^{\delta -}$O3$'$H group to a water molecule occur concurrently with the nucleophilic attack of $^{\delta -}$O3$'$ on P$^{\delta +}_{\alpha}$ catalyzed by Mg$^{2+}$ ions. Note that we have used $\delta +$ and $\delta -$ instead of 5+ and 2$-$ for the phosphorous and oxygen ions, respectively, in order to accommodate the changes of these charges as a result of their changing polarizabilities in the presence of changing interaction strengths. In fact, we will implement this notation throughout the paper for any element in any molecule without caution. The readers need not worry about this, because when the time comes, we will identify their magnitudes with a set of properly defined inequalities. Any number that carries a prime (3$'$ or 5$'$) refers to a specific carbon atom in a given molecule.  

All the computational analyses thus far focused on the ``pre- and post-chemistry avenues'', which also correspond to the close- and open-conformations of pol-$\beta$ and DNA complex~\cite{rad,alb,others,aba}. In these reports~\cite{rad,alb,others,aba}, several responsible reaction-pathways were proposed entirely based on these avenues, disregarding the biochemical reaction route (with the highest specificity and fidelity) in the chemical process itself, sandwiched between those two avenues (for more information, please refer the Section, Additional notes, given prior to conclusion). Tracking this route (within the chemical process) is far more efficient as there can only be a limited number of reaction routes to initiate the nucleophilic attack that could insert a single base to the growing DNA strand. Therefore, it is technically easier, accurate and reliable if we could exploit the information obtained from this chemical process to narrow down the processes in the pre- and post-chemistry avenues. Here we make use of the ionization energy theory~\cite{pra} and the energy-level spacing renormalization group method~\cite{aop} to evaluate the van der Waals (vdW) interaction of a stronger type, down to the electronic level. We will show that such an evaluation will lead us to find the exclusive biochemical reaction (assisted by (i) the Hermansson blue-shifting hydrogen bond~\cite{her1,her2,her3} and (ii) the Arunan composite hydrogen-vdW bond~\cite{arun,arun2,arun3,arun4} effects) with the highest specificity and fidelity, in the presence of the correct Watson-Crick hydrogen bonds~\cite{watson}. 

\section{Theoretical method}

Here, we first introduce the ionization energy theory (IET) and its approximation, which are needed to renormalize the vdW interaction, and to determine the average atomic energy-level spacing. This is followed by the introduction to the renormalized vdW interaction, which will be shown to be responsible for the (bio)chemical reactions if this vdW is of a stronger type, much stronger than the standard vdW interaction that has a $1/R^6$ dependence, where $R$ is the separation between two nuclei.

\subsection{Ionization energy approximation}

Our analyses here use the ionization energy approximation, which can be understood in a straightforward manner if we start from the IET-Schr$\ddot{\rm o}$dinger equation~\cite{pra}, 
\begin {eqnarray}
i\hbar(\partial \Psi(\textbf{r},t)/\partial t) &=& \big[-(\hbar^2/2m)\nabla^2 + V_{\rm IET}\big]\Psi(\textbf{r},t) = H_{\rm IET}\Psi(\textbf{r},t) \nonumber \\&=& (E_0 \pm \xi)\Psi(\textbf{r},t), \label{eq:IN14}
\end {eqnarray}  
where $\Psi(\textbf{r},t)$ is the time-dependent many-body wave function, $\hbar = h/2\pi$, $h$ denotes the Planck constant and $m$ is the mass of electron. The most important parameter here is the eigenvalue, $E_0 \pm \xi$, which is the real (true and unique) energy levels for a given quantum system (an atom or a molecule or a solid or anything in between). Equation~(\ref{eq:IN14}) reduces to the standard Schr$\ddot{\rm o}$dinger equation by noting that $E_0 \pm \xi = E$, which is exact by definition, and it is not trivial~\cite{pra}. Here, $E_0$ is the energy levels at zero temperature and in the absence of any external disturbances. This means that $E_0$ is a constant, while $\pm\xi$ is known as the ionization energy (or the energy-level spacing) where $\pm$ refers to electrons and holes, respectively. In this work, $-\xi$ does not play any role for an obvious reason, but holes (positive charge carriers) can play a vital role in biomolecules as proven experimentally by Fujitsuka and Majima~\cite{fuji} with respect to charge transfer in DNA. There are two formal proofs for the ionization energy approximation. The first one is indirect and reads, $H_{\rm IET}\Psi(\textbf{r},t) = (E_0 \pm \xi^{\rm quantum}_{\rm matter})\Psi(\textbf{r},t) \propto (E_0 \pm \xi^{\rm constituent}_{\rm atom})\Psi(\textbf{r},t)$, available in Ref.~\cite{pra}, while the second formal proof is direct (and is based on logic) and it is associated to the excitation probability of electrons and holes within the ionization energy based Fermi-Dirac statistics~\cite{physc,pla}. 

The ionization energy approximation reads
\begin {eqnarray}
&&\xi^{\rm quantum}_{\rm matter} \propto \xi^{\rm constituent}_{\rm atom}, \label{eq:IN14b}
\end {eqnarray}  
where $\xi^{\rm quantum}_{\rm matter}$ is the real energy level spacings of a particular quantum system. In the most general sense, $\xi$ is defined to be the real energy spacings between energy levels such that an electron from one energy level needs to overcome an energy cost when that electron tries to occupy another energy level. In molecules, $\xi$ refers to the energy level spacing between an occupied level (in the highest occupied molecular orbital (HOMO)) and an empty level (in the lowest unoccupied molecular orbital (LUMO)), which is the molecular energy level spacing. Due to the above definition for $\xi$, the quantum matter discussed in this work refers to molecules. Now, we can readily predict the changes to $\xi^{\rm quantum}_{\rm matter}$ if we know $\xi^{\rm constituent}_{\rm atom}$, where the values for $\xi^{\rm constituent}_{\rm atom}$ can be obtained from the experimental atomic spectra available in Refs.~\cite{web,web2}. For example, $\xi_{\rm H_2O} \propto [2\xi_{\rm H^+} + \xi_{\rm O^+}]$. The ionization energy approximation becomes exact if the quantum matter are atoms or ions because $\xi^{\rm quantum}_{\rm matter} = \xi^{\rm atom}_{\rm ion}$.   

In addition, we can also prove why an atomic hydrogen takes up the role of a cation in a water molecule, while an oxygen atom ends up as an anion. To prove this, we need to verify $\xi_{\rm 2H^+}$ $<$ $\xi_{\rm O^{2+}}$, which can be confirmed from Refs.~\cite{web,web2} where $\xi_{\rm 2H^+}$ (1312 kJmol$^{-1}$) $<$ $\xi_{\rm O^{2+}}$ (2351 kJmol$^{-1}$). Here, the averaging follows Eq.~(\ref{eq:24b}) (given below). This inequality also proves that the formation of O$_2$H is never possible. Therefore, an atomic hydrogen gives up (donates) its electron to an oxygen atom, and this process is never the other way round (having oxygen as a donor) because there is always a stronger Coulomb attraction between the electron from H and the oxygen nucleus. In other words, if $\xi$ is larger for atomic A compared to atomic B, then the nucleus of atomic A will always attract the valence electrons from atomic B with a stronger Coulomb force, and not \textit{vice versa}~\cite{pra}. This reinforces the idea that an atomic hydrogen acts as a cation, whereas, an oxygen atom takes up the role of an anion, in a system consists of H and O. Consequently, one can use this (ionization energy) approximation to evaluate the changing interaction strengths between two atoms or ions from different molecules. However, the bonds between atoms do not necessarily involve a single valence electron from each atom, and thus, we need to average the ionization energies for atoms that donate (or share) more than one electron. In this case, the ionization energy averaging follows 
\begin {eqnarray}
&&\xi^{\rm constituent}_{\rm atoms} = \sum_j\sum_i^z (1/z)\xi_{j,i}(\texttt{X}^{i+}_{j}), \label{eq:24b}   
\end {eqnarray}
where each subscript $j$ represents one chemical element ($\texttt{X}_j$) in a particular molecule, while $i = 1, 2, \cdots, z$, where $i$ counts the number of valence electrons coming from each chemical element. For example, if $\texttt{X}_{j = 1}$ = P such that P forms five bonds with its nearest neighbors, then P $\rightarrow$ P$^{5+}$, and therefore,  
\begin {eqnarray}
&&\xi_{\rm P^{5+}} = (1/5)\sum_i^5 \xi_{{\rm P}^{i+}} = 3414~ {\rm kJmol^{-1}}. \label{eq:24c}   
\end {eqnarray}
The first, second, $\cdots$, $z^{\rm th}$ ionization energies (prior to averaging) were obtained from Refs.~\cite{web,web2}. For a water molecule, $\texttt{X}_{j = 1}$ = H and $\texttt{X}_{j = 2}$ = O. It is to be noted here that the ionization energy theory was first formally proven to be a renormalized theory within the energy-level spacing renormalization group method~\cite{pra,aop}. This energy-level spacing renormalization procedure can be related exactly to the Shankar renormalization technique within the screened Coulomb potential derivation, and his renormalization method is documented here~\cite{shank1,shank2,shank3}. Further experimental supports for IET and its applications can be found here~\cite{pccp,cpc,dt,jap,KD1,pccc,sis,kris,radha}. 

Moreover, the importance of ionic valence states on the electronic transport properties of oxides and magnetic semiconductors was first experimentally proven by Mahendiran \textit{et al}.~\cite{mahen}, the computational (using the Density Functional Theory) proof was given by Mahadevan and Zunger~\cite{maha}, while the theoretical proofs (using IET) have been developed in Ref.~\cite{pssb}. In these reports~\cite{mahen,maha,pssb}, the valence state of any cation is shown to vary depending on its surrounding. Therefore, our averaging procedure explained earlier, properly takes the valence-state effect into account. The ``surrounding'' here refers to the nearest, and the next nearest neighbors only, such that these neighbors are directly bonded to this cation. For a given biochemical process however, the ``surrounding'' also needs to consider the neighbors that are not directly bonded to a cation or an anion. This additional interaction originating from the non-bonding neighbors can be taken into account by means of changing interaction strength between two atoms or ions (see below for details).

\subsection{Renormalized van der Waals interaction for biochemical reactions}

Earlier, we have justified why the ionization energy approximation allows us to track the interaction strength between atoms from different molecules. Here, we need to capture that an increasing attractive strength between two atoms can lead to a (bio)chemical reaction. To do that, we need to develop a new vdW formalism that allow a stronger Coulomb attraction between two atoms (or ions) for all $R \geq r_1 + r_2$ where $R$ is the separation between two nuclei, $r_1$ and $r_2$ denote the radii of two interacting atoms or ions~\cite{ion}. This stronger Coulomb attraction has been captured using the renormalized vdW formalism, and the relevant formula is~\cite{ion}      
\begin {eqnarray}
\tilde{V}'_{\rm Waals}(\xi) = V^{\rm e-ion}_{\rm Coulomb} + (1/2)\hbar\omega_0\big((1/\sqrt{2}) - 1\big)\exp{\big[(1/2)\lambda\xi\big]}, \label{eq:1.2}
\end {eqnarray}  
where, $\hbar\omega_0$ denotes the ground state energy (of a given molecule) in the absence of external disturbances, and for $T = 0$K, and $\tilde{V}'_{\rm Waals}(\xi)$ is the stronger version of the attractive vdW interaction, after renormalization~\cite{ion}. Moreover, $\tilde{V}'_{\rm Waals}(\xi)$ is negative guaranteed by the inequality $1/\sqrt{2} - 1 < 0$ and the attractive Coulomb interaction between an electron and a nucleus is also negative by definition. Here, $\hbar$ is the Planck constant divided by 2$\pi$, $\lambda = (12\pi\epsilon_0/e^2)a_{\rm B}$, in which, $a_{\rm B}$ is the Bohr radius of atomic hydrogen, $e$ and $\epsilon_0$ are the electron charge and the permittivity of free space, respectively. It is interesting to note here that Arunan \textit{et al}. were the first to point out the existence of a composite bond that consists of hydrogen bond and the vdW attraction~\cite{arun,arun2,arun3,arun4}, which has been captured by Eq.~(\ref{eq:1.2}) by means of the generalized vdW interaction.  

The detailed derivation for $\tilde{V}'_{\rm Waals}(\xi)$, its physical insights and why it is suitable for both long and short distances are given elsewhere~\cite{ion}. Equation~(\ref{eq:1.2}) has been proven to be stronger than the standard vdW prediction due to the existence of an asymmetric polarization between two atoms. The relevant equation that associates $\xi$ to the atomic displacement polarization is given by~\cite{polar} 
\begin {eqnarray}
\alpha_{\rm d} = \frac{e^2}{M}\bigg[\frac{\exp[\lambda(E_{\rm F}^0 - \xi)]}{(\omega_{\rm{ph}}^2 - \omega^2)}\bigg], \label{eq:1.2polar}
\end {eqnarray}  
where $\omega$ is the electric-field dependent frequency due to valence-electron displacement, $\omega_{\rm{ph}}$ denotes the phonon frequency of undeformable ions (mass, $M$) and $E_{\rm F}^0$ is the Fermi energy at zero Kelvin. A stronger polarization exists if the two atoms have very different ionization energy values such that $\xi^{\rm atom}_{\rm A} < \xi^{\rm atom}_{\rm B}$ where atomic A will become a cation, while atomic B assumes the role of an anion based on IET. Presently~\cite{rad,alb,others,aba,war,pan}, and usually, this asymmetric polarization are assumed to be static (by assigning fixed charges to atoms or ions) and therefore, the interaction between any two ions are electrostatic. For example, the red- and blue-shifted hydrogen bonds were determined entirely from the step-by-step changes to the distance between the supposedly hydrogen-bonded atoms~\cite{war,pan}. It is physically alright to assume the charges assigned to the atoms of a particular molecule to be static where this assumption can be used to determine the electrostatic dipole-dipole, dipole-induced dipole and/or the London dispersion forces. For example, the above stated interactions can become stronger or weaker entirely from the step-by-step changes to the separation between two strongly interacting atoms ($R$), however, this assumption (by definition) ignores the systematic changes to the atomic polarization when $R$ changes. 

The above mentioned atomic polarization and the vdW forces are often treated as adjustable parameters in the \textit{ab-initio} quantum mechanical calculations, and also in molecular dynamical simulations~\cite{gal1}. In particular, in any \textit{ab-initio} quantum mechanical calculations, polarization functions are introduced by hand to allow some variations to the atomic orbitals in order to evaluate the molecular integrals. For example, within the Gaussian basis set formalism, each basis function (indicated with a subscript ``$r$'') represents a normalized linear combination of a few Gaussians, namely, the contracted Gaussian~\cite{ira} 
\begin {eqnarray}
\chi_r = \sum_{ijk}d_{ijk;r}g_{ijk}, \label{eq:ira1}
\end {eqnarray}
where $d_{ijk;r}$ is the contraction coefficient, the primitive Gaussian based on the Cartesian coordinate (centered on atom b),  
\begin {eqnarray}
g_{ijk} = Nx_{\rm b}^iy_{\rm b}^jz_{\rm b}^k\exp({-\alpha r^2_{\rm b}}), \label{eq:ira2}
\end {eqnarray}
and the Cartesian-Gaussian normalization constant,  
\begin {eqnarray}
N = \big(2\alpha/\pi\big)^{3/4}\big[(8\alpha)^{i + j + k}i!j!k!/(2i)!(2j)!(2k)!\big]^{1/2}. \label{eq:ira3}
\end {eqnarray}
Here, Eqs.~(\ref{eq:ira1}),~(\ref{eq:ira2}) and~(\ref{eq:ira3}) are designed (not through an \textit{ab-initio} method) to allow polarizations where $i$, $j$ and $k$ are nonnegative integers, $\alpha > 0$ is an orbital exponent, and $x_{\rm b}$, $y_{\rm b}$, $z_{\rm b}$ are Cartesian coordinates such that the nucleus b is the origin. When $i + j + k = l = 0$, one has the $s$-type Gaussian, while the polarizable $p$- and $d$-type Gaussians can be obtained for $l = 1$ and $l = 2$, respectively. Equation~(\ref{eq:ira1}) represents the guessed linear combination of atomic orbitals (LCAO), which allow distortions such that an atomic orbital's (AO) shape and its center of charge can be shifted upon the formation of a molecule, by requiring $l > 0$. The point is, the above AO with $l > 0$ are introduced at the outset by hand, and these AO-functions (see Eq.~(\ref{eq:ira1})) are not determined from any \textit{ab-initio} quantum mechanical calculation or derivation~\cite{ira}, and therefore, we can now understand why the parameters associated to polarizations were treated as variationally adjustable parameters~\cite{gal1}.  

As a consequence, we need to find an alternative way to understand how and why the change in the polarization strength can influence the blue- or red shifting hydrogen bond (or the generalized vdW interaction), in the presence of another ion. These information can be extracted from the \textit{ab-initio} calculations by imposing some guessed AO-functions (Eq.~(\ref{eq:ira1})) at the outset, and then treat the parameters related to polarizations as variationally adjustable. In our approach, the polarizations are determined directly without any inbuilt guessed functions or adjustable parameters. On the other hand, if one considers an electrostatic interaction, then an induced polarization can change the charge magnitude that was originally assigned to an ion, making such an assumption inadequate to study chemical reactions. Here, we will use the ionization energy approximation and the renormalized vdW theory to tackle the polarization effect that is responsible to initiate a chemical reaction between two strongly interacting atoms (or ions) from different molecules. In the following analyses, we will show that the changes in the polarization caused by different surrounding molecules give rise to the corresponding changes to the partial charges of the reactant atoms. These changes in the partial charges assigned to reactant atoms or ions (from different molecules) can lead to a lock-in configuration, which can proceed to a chemical reaction, if the configuration has the strongest Coulomb attraction.

\subsection{Benchmark analyses: peptide bond formation (condensation reaction)} 

Here, we apply IET and the new vdW formula to evaluate the formation of a peptide bond between two $\alpha$-amino acids, 
\begin {eqnarray}
&&\rm H_2NCH\texttt{R}COOH + H_2NCH\texttt{R}COOH \rightarrow H_2NCH\texttt{R}CONHCH\texttt{R}COOH + H_2O, \label{eq:1.2b}
\end {eqnarray}  
where $\texttt{R}$ denotes an organic substituent, which is irrelevant here. The analyses presented here involve inequalities, but they are sufficient to achieve our objective to unambiguously determine the correct biochemical reaction pathway, without the need to compute any numbers (apart from the ionization energy averaging).   

\begin{figure}
\begin{center}
\scalebox{0.4}{\includegraphics{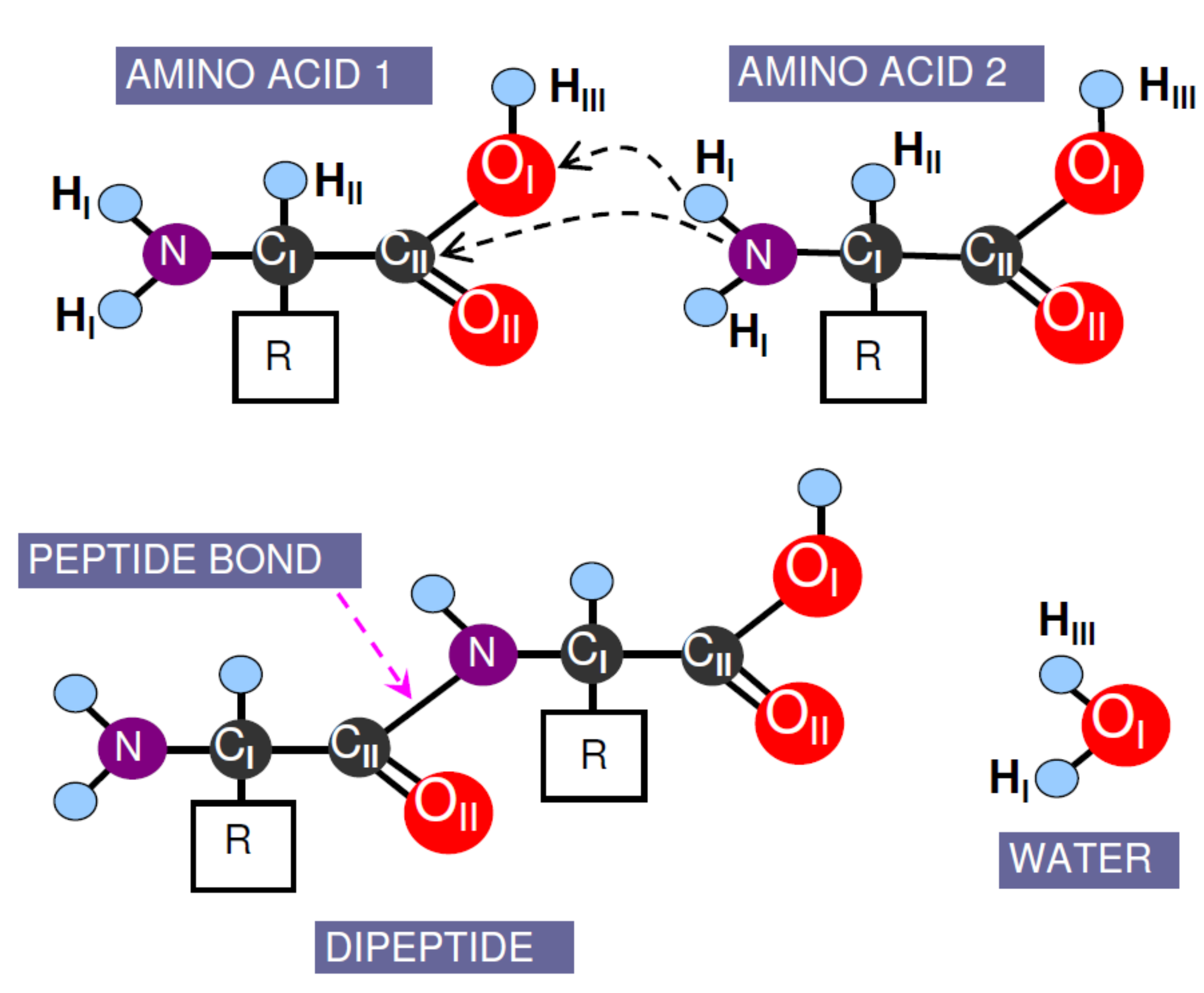}}
\caption{}
\label{fig:1}
\end{center}
\end{figure}

The condensation reaction (given in Eq.~(\ref{eq:1.2b})) is schematically depicted in Fig.~\ref{fig:1} where the dashed arrows indicate the correct reaction pathways. The alternative reaction pathways can involve C$_{\rm I}$, O$_{\rm II}$, H$_{\rm II}$ and H$_{\rm III}$ (not shown). You are kindly advised to refer to Ref.~\cite{pccp} for a much simpler system, in which the system consists of only one water molecule and a surface oxygen atom. Our strategy here is to track this condensation reaction pathway that has the strongest interaction (such that $|\tilde{V}'_{\rm Waals}(\xi)|$ from Eq.~(\ref{eq:1.2}) is maximum). This means that, a chemical reaction occurs continuously until completion or equilibrium (provided the external conditions to initiate this reaction are met), guided entirely by the changing instantaneous interaction strengths (mostly due to electron-electron and electron-ion types), not because the reactants are ``consciously aware'' that their final destination or products (after the chemical reaction) will have a smaller free-energy, or maximum entropy, or energetically more stable. There are five possibilities for two $\alpha$-amino acids to react (ignoring $\texttt{R}$), but only one of the reaction pathways is responsible for the formation of a peptide bond. We have evaluated all the reaction pathways, but we only highlight the correct reaction pathway (indicated with dashed arrows in Fig.~\ref{fig:1}) as a benchmark analysis. Both the correct and the incorrect reaction pathways for dipeptide bond formation are similar to the reaction pathways discussed for pol-$\beta$ DNA replication (given in the following section).  
   
The polarized charges ($\delta +$ or $\delta -$) for the relevant atoms in an isolated amino acid can be determined using Eqs.~(\ref{eq:IN14b}) and~(\ref{eq:24b}), while Table~\ref{Table:I} lists all the ionization-energy values required for our analyses. From these two inequalities, $\xi_{\rm N^{2+}}$(2129 kJmol$^{-1}$) $>$ $\xi_{\rm 2H^{+}}$(1312 kJmol$^{-1}$) and $\xi_{\rm N^{4+}}$(4078 kJmol$^{-1}$) $>$ $\xi_{\rm C^{4+}}$(3571 kJmol$^{-1}$), we can conclude that N is negatively charged, N$^{\delta -}$, H$_{\rm I}$ is positively charged, H$_{\rm I}^{\delta +}$, while C$_{\rm I}$ is slightly negatively charged, C$_{\rm I}^{\delta -}$ because even though H$_{\rm II}$ contributes an electron to C$_{\rm I}$, the negative charge due to this electron (from H$_{\rm II}$) is compensated by the electron (from C$_{\rm I}$) that is polarized toward N to create a covalent bond with N. This electron from C$_{\rm I}$ is actually asymmetrically polarized toward N due to $\xi_{\rm N^{4+}}$(4078 kJmol$^{-1}$) $>$ $\xi_{\rm C^{4+}}$(3571 kJmol$^{-1}$). Therefore, $^{\rm H_{\rm I}}\delta +$ $>$ $^{\rm H_{\rm II}}\delta +$ and $^{\rm N}\delta -$ $>$ $^{\rm C_{\rm I}}\delta -$. In addition, a C---C bond gives rise to an isotropic polarization (because $\xi_{\rm C} = \xi_{\rm C}$), and thus, C does not gain any charge ($\delta +$ or $\delta -$) by making bonds with another C. However, C$_{\rm II}^{\delta +}$ is positively charged because $\xi_{\rm C^{4+}}$(3571 kJmol$^{-1}$) $<$ $\xi_{\rm O^{4+}}$(4368 kJmol$^{-1})$, and no charge contribution from C$_{\rm I}$ to C$_{\rm II}$ due to isotropic polarization. Therefore, we have these inequalities, $^{\rm C_{\rm II}}\delta +$ $>$ $^{\rm H_{\rm I}}\delta +$ $>$ $^{\rm H_{\rm II}}\delta +$ and $^{\rm O_{\rm I}}\delta -$ $>$ $^{\rm O_{\rm II}}\delta -$. The second inequality is due to $\xi_{\rm 4H^{+}}$(1312 kJmol$^{-1}$) $<$ $\xi_{\rm C^{4+}}$(3571 kJmol$^{-1}$). Recall here that atoms with a lower averaged ionization energy easily gives away its electron, compared to an atom with a larger averaged ionization energy. 

We can now proceed to evaluate the correct (experimentally proven) reaction pathway, NH$_{\rm I}$(from amino acid 2) interacts with O$_{\rm I}$H$_{\rm III}$(from amino acid 1). H$^{\delta +}_{\rm I}$ first interacts with O$^{\delta -}_{\rm I}$, which further polarizes the O$^{\delta -}_{\rm I}$ electrons (contributed by C$^{\delta +}_{\rm II}$ and H$^{\delta +}_{\rm III}$) much closer to O$^{\delta -}_{\rm I}$. Moreover, the electron contributed by H$_{\rm I}^{\delta +}$ to N$^{\delta -}$ is now more polarized closer to N$^{\delta -}$ because H$_{\rm I}^{\delta +}$ now attracts an electron from O$^{\delta -}_{\rm I}$ originally contributed by C$^{\delta +}_{\rm II}$. These induced polarizations imply that $^{\rm H_{\rm I}\cdots O_{\rm I}-C_{\rm II}}\delta' +$ $>$ $^{\rm O_{\rm I}-C_{\rm II}}\delta +$ and $^{\rm O_{\rm I}\cdots H_{\rm I}-N}\delta'' -$ $>$ $^{\rm H_{\rm I}-N}\delta -$, where H$_{\rm I}\cdots$O$_{\rm I}$ denotes the vdW attraction within the generalized vdW formalism. Within the hydrogen-bond formalism, it is known as the blue-shifting hydrogen bond effect between H$_{\rm I}$ and O$_{\rm I}$. Here, the superscript, C$_{\rm II}-$O$_{\rm I}$ represents the covalent bond between C$_{\rm II}$ and O$_{\rm I}$. From these inequalities, $^{\rm H_{\rm I}\cdots O_{\rm I}-C_{\rm II}}\delta' +$ $>$ $^{\rm O_{\rm I}-C_{\rm II}}\delta +$ and $^{\rm O_{\rm I}\cdots H_{\rm I}-N}\delta'' -$ $>$ $^{\rm H_{\rm I}-N}\delta -$, we can now deduce that N$^{\delta''-}$ interacts more strongly with C$^{\delta' +}_{\rm II}$ thanks to the interaction between H$^{\delta +}_{\rm I}$ and O$^{\delta -}_{\rm I}$. Next we show that the above charge inequalities (due to induced polarizations) will lead us to a lock-in configuration, which sets the stage for chemical reactions.  

The lock-in configuration is obtained when H$_{\rm I}^{\delta +}$ interacts with O$^{\delta -}_{\rm I}$ and when N$^{\delta''-}$ interacts with C$^{\delta' +}_{\rm II}$ where the strength of these interactions follow Eq.~(\ref{eq:1.2}). We can now use the ionization energy approximation to calculate the relevant values, namely, $\xi_{\rm N\cdots C_{\rm II}} \propto |\xi_{\rm N^+} - \xi_{\rm C^+}|$ = 317 kJmol$^{-1}$ and $\xi_{\rm H_{\rm I}\cdots O_{\rm I}} \propto |\xi_{\rm H^+} - \xi_{\rm O^+}|$ = 2 kJmol$^{-1}$, and therefore, the $\xi_\textbf{\rm total}$ is 317 + 2 = 319 kJmol$^{-1}$, which is the highest compared to any other alternative combinations of atoms shown in Fig.~\ref{fig:1}. This highest value also gives the highest interaction strength when it is substituted into Eq.~(\ref{eq:1.2}).  

But this is only one-half of the requirement for the condensation reaction to proceed. The other half comes from the stronger bonds between C$^{\delta' +}_{\rm II}$ and O$_{\rm I}^{\delta -}$ and between N$^{\delta'' -}$ and H$_{\rm I}^{\delta +}$. The bond between C$^{\delta' +}_{\rm II}$ and O$_{\rm I}^{\delta -}$ is stronger because the electron from C$^{\delta' +}_{\rm II}$ is polarized closer to O$^{\delta -}_{\rm I}$ in the presence of H$_{\rm I}^{\delta +}$. Meanwhile, the N$^{\delta'' -}$---H$_{\rm I}^{\delta +}$ bond has also become stronger because the electron from H$^{\delta +}_{\rm I}$ is now polarized much closer to N$^{\delta'' -}$ in the presence of O$^{\delta -}_{\rm I}$. In other words, these two bonds have become stronger because they are in certain polarized modes, which are actually best suited to initiate selective chemical reactions compared to unpolarized bonds. For example, it is relatively easy to break a polarized bond via an electron and/or a proton transfer. On the contrary, unpolarized bonding electrons are more homogeneously distributed in a given molecule, and therefore, the charges for atoms within the molecule ($\delta +$ and $\delta -$) are either too small or negligible to initiate a proper and strong lock-in configuration. In the polarized modes however, atoms (in a molecule) have large polarized charges ($\delta +$ or $\delta -$), which can strongly interact with other atoms or ions from another molecule. 

Note here that thermal energy is not useful to set up a lock-in configuration because thermal energy causes non-selective electron polarizations where it will excite all the bonding electrons homogeneously in a given molecule. Hence, thermal energy can be a useful source of energy required for continuous reactions and/or to keep the activation energy lower so that a given reaction (in certain polarized modes) can proceed. 

We have identified these two bonds, C$^{\delta' +}_{\rm II}$---O$_{\rm I}^{\delta -}$ and N$^{\delta'' -}$---H$_{\rm I}^{\delta +}$ are stronger as a result of selective atomic polarizations. In other words, these bonds are more difficult to be broken by applying thermal energy, but they are more easily broken if we could lure these charged atoms with opposite charges (with an electron or a proton). In particular, the above stronger bonds will lead C$^{\delta' +}_{\rm II}$ to strongly interact with N$^{\delta'' -}$, while O$^{\delta -}_{\rm I}$ is strongly interacting with H$_{\rm I}^{\delta +}$. All we need now are some external energy supply (thermal energy for example) to initiate the simultaneous electron and proton transfers, namely an electron-transfer from N$^{\delta'' -}$ to C$^{\delta' +}_{\rm II}$ and a proton-transfer, H$_{\rm I}^{\delta +}$ to O$^{\delta -}_{\rm I}$. These transfers are likely to proceed compared to other combinations of atoms, which have been predicted from the ionization energy approximation presented earlier. This lock-in configuration has the strongest attraction, which requires the minimal external energy supply (due to the lowest activation energy). The simultaneous electron and proton transfers will complete the condensation reaction, giving rise to a dipeptide and a water molecule as reaction products. 

In summary, the charge ($\delta +$ and $\delta -$) redistribution (due to polarization) in atoms have given rise to a lock-in configuration, and subsequently have led to simultaneous electron and proton transfers in such a way that a new peptide bond is formed between N$^{\delta'' -}$ and C$^{\delta' +}_{\rm II}$, while the N$^{\delta'' -}$---H$_{\rm I}^{\delta +}$ covalent bond is broken. On the other hand, a new O$_{\rm I}^{\delta -}$---H$^{\delta +}_{\rm I}$ bond is formed, while the C$^{\delta' +}_{\rm II}$---O$_{\rm I}^{\delta -}$ bond is broken. We can now move on and use the acquired knowledge to tackle the pol-$\beta$ DNA replication.

\section{Results and discussion}

Schlick \textit{et al}.~\cite{rad,alb} found the nucleophilic attack, A1 runs simultaneously with A2 to B3 proton transfers. In contrast, we show here with unequivocal theoretical analyses that a direct deprotonation of H$^+$ from O3$'$H to O3$_{\alpha}$ (A3$_{\alpha}$ pathway) and the A1 pathway are activated simultaneously. This concurrent process (with steps A1 and A3$_{\alpha}$) is chemically driven (initiated by a stronger vdW interaction) with the highest specificity and fidelity (see Fig.~\ref{fig:2}). This direct deprotonation is also somewhat similar to Abashkin, Erickson and Burt~\cite{aba}, in which, the difference being their~\cite{aba} direct proton transfer is from O3$'$H to O2$_{\alpha}$, and then to O3$_{\alpha}$ (after bond-rotation). We use the blue-shifting vdW interaction between electrons from different molecules~\cite{ion}, or from different sections of a bigger molecule, as a determining factor for biochemical reactions, which completes one base-pair (CG) to the growing DNA strand. The letters, A, G, C and T denote the well-known DNA purine (adenine and guanine) and pyrimidine (cytosine and thymine) bases. To avoid any confusion, we will always write the name in full to denote the element carbon. 

Of course, the steps, A1 and A3$_{\alpha}$ can only be efficiently initiated in the presence of the correct (a) conformation of the incoming dCTP by interacting with the pol-$\beta$ residues and with the guanine (from the template DNA strand) via the Watson-Crick hydrogen bonds, (b) position of Mg$^{2+}$ ions and water molecules, and (c) position of pol-$\beta$ residues and their interactions with cations, anions and water molecules. In fact, these points (a to c) are valid because numerous interactions (both quantum and classical types) will come into play in a biologically dynamical setting, which have been discussed explicitly by Spyrakis \textit{et al}.~\cite{spy} and Jayaram \textit{et al}.~\cite{jaya}. Apart from those work, the readers are also referred to an excellent review by Joyce~\cite{joy} for a list of the techniques available to study biochemical reactions with respect to DNA polymerases.  

The fundamental information on the generalized vdW interaction leading to the insertion of a single base (C) to the growing DNA strand are equally (if not more) important (i) to theoretically prove the possibility of replacing phosphorous with arsenic in the Californian bacterium, strain GFAJ-1, which was experimentally observed by Wolfe-Simon and her assemblage of scholars~\cite{wolf1,wolf2}, (ii) to evaluate the \textit{in-vitro} characterization of a DNA nanomachine that maps the changes in the spatial and temporal pH, as reported by Krishnan and her colleagues~\cite{mod}, (iii) to understand how a female sex hormone (progesterone) chemically activates sperms into ``hyperactivity'' by increasing the Ca$^{2+}$ ions concentration (inside sperms)~\cite{stru,lish}, (iv) to theoretically evaluate the possibility of exploiting G-quadruplexes (a four-stranded guanine structure) as a therapeutic target in oncology to develop anticancer drugs~\cite{bala,bala2}, and (v) to study the initiation of human colon tumor that grows like a cancer due to pol-$\beta$ DNA replication error~\cite{muni}.

\subsection{Biochemical reactions in the chemical stage}

In this work, we track the electron-electron and electron-ion interaction strengths via the changing energy-level spacing (presented earlier in the theoretical method section) during a particular biochemical reaction. Interestingly, the band structure and energy levels of proteins have also been suggested to be useful to understand proteins electronic interaction~\cite{evans}. Here, we will track the above interaction strengths so as to understand why and how the nucleophilic attack proceeds with a unique path. In fact, there is only one approach that is well-defined and allows one to expose the chain of events leading to this exclusive path occurring at the critical point, which was first introduced in Ref.~\cite{pccp} to capture the origin of a single water-molecule splitting on a MgO surface. The formal proofs that justify this approach are given elsewhere~\cite{prsa,qat}. Add to that, these interaction strengths that we will be tracking are directly related to the ionization energy theory (IET) and the renormalization group method, which are also mathematically well-defined~\cite{pra,aop}. 

Furthermore, our analyses can be made quantitative by invoking the ionization energy approximation, which gives us the license to claim that the molecular energy level spacings are directly proportional to the atomic energy level spacings of all the atom exist in a given molecule~\cite{pla,pla2}. The atomic ionization energies of all the atoms in a molecule is of course different from the molecular ionization energies of that particular molecule. They are different because the atomic energy levels will have to transform (due to wave function transformation) during a chemical reaction to form the molecular energy levels~\cite{prsa,qat}. However, the energy level spacings of a molecule containing certain atoms has been proven to be proportional (not equal) to the averaged atomic energy level spacings of all the isolated atoms present in that particular molecule~\cite{pra}.     

What we need to do now is to first understand the chain of events giving rise to the validity of the steps, A1 and A2 within IET. When the negatively charged $^{\delta -}$O3$'$ comes closer to P$_{\alpha}$ (see Fig.~\ref{fig:2}), the $^{\delta -}$O3$'$---H$^{\delta +}$ bond is further reinforced, which is the effect we discovered in Ref.~\cite{pccp}. The reason is that the electrons contributed by H and carbon to create the O3$'$---H and O3$'$---carbon covalent bonds are now being strongly bound to O3$'$ in the presence of P$^{\delta +}_{\alpha}$, which pulls these electrons further away from carbon and H (from now on, we denote this specific hydrogen as H$'$). The $^{\delta -}$O3$'$---$^{\delta +}$H$'$ covalent bond is necessarily stronger due to induced electrostatic interaction between $^{\delta -}$O3$'$ and P$_{\alpha}$. The reason is that the bonding electron (contributed by H$'$) is now more polarized toward $^{\delta -}$O3$'$ due to P$_{\alpha}$, and this will necessarily result in a stronger $^{\delta -}$O3$'$---$^{\delta +}$H$'$ covalent bond. Logical proof: the above covalent bond has to get stronger, or weaker, and definitely, its strength cannot remain the same in the presence of P$^{\delta +}_{\alpha}$. Here, all other conditions from the pre-chemistry avenue are relatively weak. They are weak because in the chemical stage, these external effects coming from the pre-chemistry avenue do not directly involve in the chemical reaction. Of course, the molecules, ions and the residues from the pre-chemistry avenue do set the stage for the chemical reaction to proceed, but they do not directly involve in the nucleophilic attack as reactants. Now, let us assume the said bond ($^{\delta -}$O3$'$---$^{\delta +}$H$'$) gets weaker in the presence of P$^{\delta +}_{\alpha}$, and this is only possible if the bonding electron from H$'$ is weakly polarized toward O3$'$. Therefore, one can immediately deduce that this leads to a contradiction because the bonding electron contributed by H$'$, has to be more polarized toward O3$'$ due to the attractive induced interaction between P$^{\delta +}_{\alpha}$ and $^{\delta -}$O3$'$.   

Consequently, it is energetically inconvenient to break this stronger O3$'$---H$'$ covalent bond by brute force (by thermal energy), such as the electronic excitation of the (O3$'$---H) bonding electron, which is now strongly polarized toward P$^{\delta +}_{\alpha}$. Therefore, it would be easier by allowing the oxygen (O$^{\delta -}_{\rm WAT}$) from a nearby water molecule to form the required hydrogen bond (between O$_{\rm WAT}$ and H$'$) that can easily blue shift (\textit{via} the Hermansson blue-shifting hydrogen bond effect) due to an attractive interaction between O3$'$ and P$_{\alpha}$. Let us rephrase the last sentence due to its extreme importance and also because of its far-reaching effect in biological systems. The stronger covalent bond between O3$'$ and H$'$ is due to the attractive interaction between O3$'$ and P$_{\alpha}$, and is actually more easily breakable, not by brute force, but when it is subjected to an induced electrostatic interaction provided by the O$^{\delta -}_{\rm WAT}$ from a water molecule. This means that, the original covalent bond between O3$'$ and H$'$ is not easily breakable by a water molecule in the absence of the attractive interaction between O3$'$ and P$_{\alpha}$. The readers are kindly advised to remember this point, or recall the theoretical method section as we will not stress it again when used later.

Now, back to Hermansson, we know that her blue-shifting interaction can initiate the Radhakrishnan-Schlick deprotonation (step A2). Unfortunately, due to this deprotonation (losing H$'$ to a water molecule), one now requires some external energy from an unknown source (this could be some additional heat or interaction) to overcome the activation energy and initiate the nucleophilic attack (A1). In other words, by having H$'$ bonded to O3$'$, one has the strongest interaction between the bonding electron (coming from O3$'$---H$'$) and the nucleus of P$^{\delta +}_{\alpha}$ leading to smaller activation energy. To see why and how this activation energy is smaller, we just need to wait a bit longer because we need to introduce and build the arguments properly. Anyway, the other subsequent proton transfers will not be captured with IET because these latter paths, B1 to B3 are similar to the hydrogen-bonding mechanism stated above, and they are not unique. They are not unique because even though these steps, A1, A2 and B1 to B3 require minimal total energy compared to other pathways (that involve different residues and so on), one can always manipulate any number of pathways to further reduce the minimal total energy. For example, we can have a direct deprotonation from the H$_2$O- - -H$'$ to one of the oxygen in the outgoing pyrophosphate without involving D190, by manipulating the ``constrained'' parameters. Here, the dashed lines, ``- - -'' means a hydrogen bond, ``---'' denotes a covalent bond, and ``='' is a double covalent bond used in text. 

In contrast, we prove here the existence of a unique reaction pathway that consists of two simultaneous steps, (A1) $^{\delta -}$O3$'$ attacks P$^{\delta +}_{\alpha}$ while (A3$_{\alpha}$) $^{\delta -}$O3$_{\alpha}$ counterattacks $^{\delta +}$H$'$ (from O3$'$) that has the highest specificity and fidelity. We will also explain why and how these steps are strategically interdependent that make them exclusive. But before we dig deep, let us first properly introduce the ionization energy approximation relevant to this work. The origin of these charges, $\delta -$ and $\delta +$ are due to their respective polarization, which can be understood from the average atomic ionization energy ($\xi$) values~\cite{cpc}. Here, large ionization energy value implies large energy needed to polarize or to excite the electron from its initial or ground state. For example, one can obtain the average ionization energy values from Table~\ref{Table:I}~\cite{web} and write $\xi_{2\rm H^{+}}(1312$~kJmol$^{-1}$) $< \xi_{\rm O^{2+}}(2351$~kJmol$^{-1}$), and therefore, H and O are indeed positively and negatively charged, as they should be. This inequality proves that oxygen can never be a cation in water molecules. Using the same arguments, we can obtain the inequality, $\xi_{\rm P}$ $< \xi_{\rm O}$, which is also true for all individual (1$^{\rm st}$, $\ldots$, 5$^{\rm th}$) ionization energies (see Table~\ref{Table:I}) to justify the type of charges ($\delta +$ or $\delta -$) carried by these elements, P$^{\delta +}_{\alpha}$ and O$^{\delta -}$ in the 5$'$-triphosphate group.

Up to this point, it is true that we have no constrained parameters to handle because we deliberately ignored them, not because ignorance is bliss, but to keep all the analyses exact and unambiguous within the chemical process, sandwiched between the pre- and post-chemistry avenues. The complex processes and their numerous possibilities involved in these avenues necessitate the introduction of constrained parameters. To avoid such parameters, the chemical process is isolated, which is theoretically valid because the interactions during this closing and opening stages, do not directly involve in the biochemical reactions that insert a base to the growing DNA strand. Thus, in the following paragraphs, our strategy is to first implement this isolation correctly, and then discuss the possibility of carrying out the pre- and post-chemistry investigations subject to the exclusive biochemical reaction pathway.

Apart from the well-known nucleophilic attack (A1 pathway), there are four other possible chemical reactions that correspond to four oxygen atoms bonded to P$_{\alpha}$, namely, P$_{\alpha}$---O3$_{\alpha}$---P, P$_{\alpha}$$=$O2$_{\alpha}$, P$_{\alpha}$---O---carbon, and P$_{\alpha}$---O$^{-}$ (see Fig.~\ref{fig:3}A to D), in which only one of them should occur simultaneously (with A1). Apparently, we need to eliminate any three of them by considering each in turn, and then prove A3$_{\alpha}$ has the highest specificity and fidelity. We have to warn the readers that the following analyses are entirely analytic and are inevitably subtle, plagued by different magnitudes of polarization for the element, P$_{\alpha}$, and the four oxygen atoms bonded to P$_{\alpha}$ in the presence of O3$'$, H$'$, P (valid only for the oxygen in P$_{\alpha}$---O3$_{\alpha}$---P) and carbon (valid only for the oxygen in P$_{\alpha}$---O---carbon).

From the ionization energy approximation (given earlier), we can understand the validity of the assigned charges in P$^{\delta +}_{\alpha}$---O$^{\delta -}$---carbon$^{\delta +}$, and consequently the steps, A1 and A3$_{\alpha}$ are also correct from the induced Coulomb interaction between $\delta +$ and $\delta -$. Of course, this induced interaction has a deep microscopic origin, which will activate the required chemical reactions to extend the DNA strand. But it is sufficient to consider only the partial charges now (without going into the details), to rule out the three out of the four steps listed above, and to find the crucial oxygen atom (bonded to P$_{\alpha}$) that interact with the strongest induced interaction with H$'$. Once the correct oxygen is identified, we can then go on and exploit the existence of the renormalized vdW interaction of a stronger type (Eq.~(\ref{eq:1.2})) that is responsible for the simultaneous nucleophilic attack A1, and the counterattack A3$_{\alpha}$. The nucleophilic- and the counter-attack depicted in Fig.~\ref{fig:3}A to D are due to an induced attractive electrostatic interaction between $^{\delta -}$O3$'$ and P$^{\delta +}_{\alpha}$. This induced interaction is a prerequisite to form a new covalent bond ($^{\delta -}$O3$'$---P$^{\delta +}_{\alpha}$), which can be obtained from the stronger vdW interaction (see Eq.~(\ref{eq:1.2})). Using Eq.~(\ref{eq:1.2}), we can go on to untangle each chemical step (sandwiched between the pre- and post-chemistry avenues) involved during the base, C insertion (from dCTP), and find the crucial oxygen (bonded to P$_{\alpha}$) that strongly interact with H$'$ during the nucleophilic attack. 

\subsection{A1 and A3$_{\alpha}$ steps}

\begin{figure}
\begin{center}
\scalebox{0.3}{\includegraphics{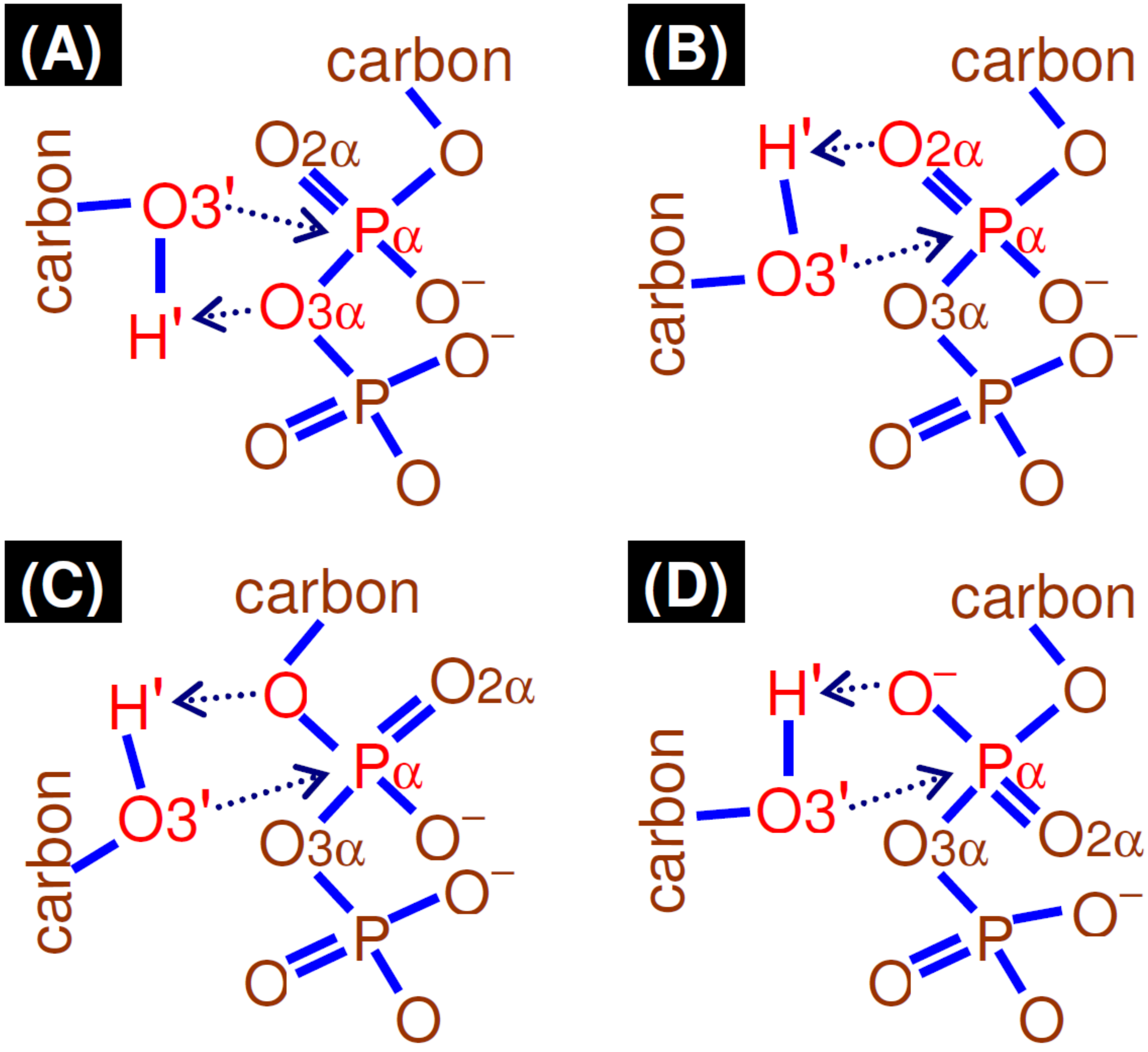}}
\caption{}
\label{fig:3}
\end{center}
\end{figure}

The moment $^{\delta -}$O3$'$ unleashes its polarizable electron on P$^{\delta +}_{\alpha}$, $^{\delta -}$O3$_{\alpha}$ strikes back instantaneously, attacking $^{\delta +}$H$'$. As a matter of fact, these nucleophilic- and counter-attacks are simultaneous to the extent that one can also view it in the reverse attacking order. But this is incorrect. The above original order-of-attack (nucleophilic-attack invites the counterattack) can be understood by noting $^{\rm P_{\alpha}}$$\delta +$ $>$ $^{\rm H'}$$\delta +$ and $^{\rm O3'}$$\delta -$ $>$ $^{\rm O3_{\alpha}}$$\delta -$, which imply a stronger attractive interaction between $^{\delta -}$O3$'$ and P$^{\delta +}_{\alpha}$. The first inequality is obvious with four oxygen atoms (forming five covalent bonds) with P$_{\alpha}$. Whereas, the latter inequality is due to $\frac{1}{2}[\xi(\rm P^{5+}_{\alpha}) + \xi(\rm P^{5+})]$ = 3414 kJmol$^{-1}$, $\frac{1}{2}[\xi(\rm carbon^{4+}) + \xi(\rm H^{+})]$ = 2442 kJmol$^{-1}$, and therefore, $\frac{1}{2}[\xi(\rm P^{5+}_{\alpha}) + \xi(\rm P^{5+})] > \frac{1}{2}[\xi(\rm carbon^{4+}) + \xi(\rm H^{+})]$. Here, the averaging follows Eq.~(\ref{eq:24b}). In other words, the electrons contributed by carbon and H are of lower ionization energies, and therefore more strongly polarized toward $^{\delta -}$O3$'$. While the electrons responsible for the partial charge in $^{\delta -}$O3$_{\alpha}$ are coming from the less polarizable P (see Fig.~\ref{fig:3}A). Here, P is the least polarizable element (with five bonds) compared to carbon (with four bonds) and H (with fixed one bond). All bonds are predominantly covalent-type, unless stated otherwise.

The chemical steps depicted in Fig.~\ref{fig:3}A are also strategically interdependent. (i) The nucleophilic attack will first reinforce the O3$'$---H$'$ bond, however, this stronger bond will be easily breakable only if H$'$ is subjected to $^{\delta -}$O3$_{\alpha}$. (ii) Conversely, the counterattack will only make the P$_{\alpha}$---O3$_{\alpha}$ bond stronger, but again, this bond is also easily breakable if P$^{\delta +}_{\alpha}$ is allowed to interact with $^{\delta -}$O3$'$. As a consequence, we can observe that the two steps, (i) and (ii) nicely correspond to A1 and A3$_{\alpha}$, and they are precisely interdependent. For example, both (i) and (ii) when proceed simultaneously, cause these bonds, P$_{\alpha}$---O3$_{\alpha}$ and O3$'$---H$'$ to break, which add C to the growing DNA strand due to the formation of P$_{\alpha}$---O3$'$ bond. On the other hand, a PP$_i$ molecule leaves the scene after the formation and breaking of O3$_{\alpha}$---H$'$ and P$_{\alpha}$---O3$_{\alpha}$ bonds, respectively. Recall here that we follow the path that likely to have the strongest electron-electron and electron-nucleus interactions in accordance with Ref.~\cite{ion}. These strongest interactions will require minimal activation energy for a particular reaction to proceed (hence, require smaller external energy supply).

It is clear that the process consisting of A1 and A3$_{\alpha}$ steps is a type of lock-in process due to strong interdependence that can be made exclusive with the highest specificity and fidelity, if only we could systematically rule out the other oxygen atoms bonded to P$_{\alpha}$. This interdependence is actually initiated by a stronger vdW attraction between O3$'$ and P$_{\alpha}$, and between O3$_{\alpha}$ and H$'$ with further assistance from the Hermansson blue-shifting hydrogen bond effect. The formation of these bonds, P$_{\alpha}$---O3$'$ and H$'$---O3$_{\alpha}$ are assisted by the Hermansson blue-shifting hydrogen bond effect, at least for the bond, H$'$---O3$_{\alpha}$. Whereas, the formation of P$_{\alpha}$---O3$'$ bond is due to the Hermansson-like blue-shifting hydrogen bond effect. The readers are referred to Ref.~\cite{pccp} for some straightforward details on the applications of the Hermansson-like bonds. Consequently, these steps do not require large external energy supply to break the above-stated bonds.  

The process shown in Fig.~\ref{fig:3}B consists of the Abashkin-Erickson-Burt deprotonation steps~\cite{aba}. In fact, the O2$_{\alpha}$ is the only close contender against the A3$_{\alpha}$ step explained above. The other oxygen atoms can be ruled out quite easily. Recall that the nucleophilic attack (step (i)) introduced earlier remains the same here, and will remain so for all other oxygen atoms (see Fig.~\ref{fig:3}C and D). The instantaneous counterattack here is between O2$_{\alpha}$ and H$'$, which will strengthen the O2$_{\alpha}$=P$_{\alpha}$ double bond that should be easily breakable only if P$^{\delta +}_{\alpha}$ is subjected to $^{\delta -}$O3$'$. There is no difference in the partial charge between $^{\delta -}$O2$_{\alpha}$ and $^{\delta -}$O3$_{\alpha}$ ($^{\rm O3_{\alpha}}\delta -$ = $^{\rm O2_{\alpha}}\delta -$) because both electrons in each case are contributed by P. However, O2$_{\alpha}$=P$_{\alpha}$ is a double bond, hence the single P$_{\alpha}$---O3$_{\alpha}$ bond is in fact twice more easily breakable than the stated double bond. Unlike the breaking of P$_{\alpha}$---O3$_{\alpha}$ bond in the A3$_{\alpha}$ step, the interaction energy from the polarized O3$'$ (during the nucleophilic attack given in Fig.~\ref{fig:3}B) is distributed equally to both bonds, until one of the double bond is broken. In contrast, for the A3$_{\alpha}$ step, the same interaction energy is fully available to break the P$_{\alpha}$---O3$_{\alpha}$ bond alone, reinforced by the counterattack between O3$_{\alpha}$ and H$'$. Thus, the process indicated in Fig.~\ref{fig:3}B is inferior to A1 and A3$_{\alpha}$ steps in terms of the highest specificity and fidelity.

In Fig.~\ref{fig:3}C, we require the O bonded to carbon and P$_{\alpha}$ to counterattack H$'$. In this case, the partial charge in the $^{\delta -}$O atom are contributed by carbon and P, whereas two P contribute two of their electrons to $^{\delta -}$O3$_{\alpha}$, implying $^{\rm O}\delta -$ $>$ $^{\rm O3_{\alpha}}\delta -$ (recall the averaging presented earlier with carbon contributing a more polarizable electron, compared to P). Due to this inequality, the O---carbon bond is the one that tend to blue-shift more than the required O---P$_{\alpha}$ bond. Therefore, the O---P$_{\alpha}$ bond cannot take advantage of the counterattack between $^{\delta -}$O and $^{\delta +}$H$'$ when P$^{\delta +}_{\alpha}$ is subjected to $^{\delta -}$O3$'$ (because the O---carbon bond will be fortified, instead of the O---P$_{\alpha}$ bond). As such, this third process can also be ruled out as it clearly lacks the required specificity and fidelity.

The logic required to exclude the final oxygen (see Fig.~\ref{fig:3}D) from the list is the easiest of them all. Here, O$^-$ counterattacks H$'$ without any disturbance to the P$^{\delta +}_{\alpha}$---O$^-$ bond. Consequently, we can immediately see that the nucleophilic attack cannot proceed until the formation of the P$_{\alpha}$---O3$'$ bond simply because this would require an additional core electron from P$_{\alpha}$ with very large $\xi$ (see Table~\ref{Table:I}) to form the 6$^{\rm th}$ bond between P$^{\delta +}_{\alpha}$ and $^{\delta -}$O3$'$. This is definitely unlikely, and so is the process suggested in Fig.~\ref{fig:3}D. 

In summary, it is an unequivocal claim that the steps, A1 and A3$_{\alpha}$ have the highest specificity and fidelity within the biochemical reaction stage sandwiched between the pre- and post-chemistry avenues, which has been pointed out much earlier. Having said that, the next stage of research can be directed to fine-tune the constrained parameters to include the conformational changes to the dCTP in such a way that the steps, A1 and A3$_{\alpha}$ can be allowed to occur. Such steps may need some three-dimensional conformational changes to the dCTP, but without disturbing the Watson-Crick hydrogen bonds, where these hydrogen bonds will remain intact as one can confirm it from Fig.~\ref{fig:2}. 

\subsection{Additional notes}

We have ignored the effects coming from the pre-chemistry avenue, for example, due to Mg$^{2+}$ ions, pol-$\beta$ residues (D190 and probably others), and water molecules. They do contribute directly in the pre- and post-chemistry avenues, but they do not act as reactants in the chemical stage. Hence, in the chemical stage, we can indeed ignore the contribution made by these molecules and ions. In fact, we can logically rule out the effects coming from the pre-chemistry avenue---any ions or charged atoms (attached to nearby molecules or residues), which are not reactants, will affect all the nearby reactant atoms (the correct as well as the incorrect ones) equally \textit{via} the electrostatic interaction. This means that, non-reactant molecules do not discriminate (not selective) when they interact \textit{via} the Coulomb interaction (be it repulsive or attractive) with the reactant atoms (directly or indirectly). Interestingly, the effects coming from the pre-chemistry avenue have also been ruled out in Ref.~\cite{war} using some molecular dynamical simulations. The results reported in Refs.~\cite{rad,alb} emphasizes the importance of pre-chemistry effects, but ignore the contribution of different oxygen atoms bonded to P$_{\alpha}$ in the chemical stage because there is no counter-attack in their calculations. This means that, their calculations in the chemical stage do not consider the interactions between H$'$ and other oxygen atoms bonded to P$_{\alpha}$ because H$'$ is lost to a water molecule.   

There are two ways to include the effect from the environment---(i) the ions and molecules from the environment affect the relevant polarization of the atoms from the reactant molecules (such that there is no chemical reaction between them) before and/or during and/or after the DNA replication, or (ii) the ions or molecules from the environment chemically react with the reactant molecules before and/or during and/or after the DNA replication. Obviously, our work ignores point (ii), but satisfies point (i). For example, point (i) can be shown to reinforce the reaction pathway (that consists of two steps, A1 and A3$_{\alpha}$) proposed herein. In particular, the cations (Mg$^{2+}$) enhance the magnitudes of the electron polarization of all anions in the vicinity (namely, the oxygen from O3$'$H and the oxygen from O3$_{\alpha}$, see Fig.~\ref{fig:2}), and these non-discriminative polarization only enhance the lock-in configuration discussed above. Here, Mg$^{2+}$ and water molecules do not selectively polarize the oxygen from O3$'$H and ignores the oxygen from O3$_{\alpha}$. This also means that, we did not constrain the position of Mg$^{2+}$ ions in such a way to obtain stronger interaction strength. Point (ii) has been considered in Refs.~\cite{rad,alb,war}. Of course, our reaction pathway becomes invalid if there is any chemical reaction between a reactant molecule and a molecule from the environment. 

Activation energy is the minimum energy needed to initiate a chemical reaction, in which the activation energy exists due to small interaction strength between two reactant species. This means that, we need to supply external energy to overcome the activation energy or microscopically, we need to increase the interaction strength between two reactant species to initiate the reaction. Here, the reaction energy is the energy needed to complete a reaction (after overcoming the activation energy). The reaction pathway (see the steps, A1 and A3$_{\alpha}$ in Fig.~\ref{fig:2}) responsible for the DNA replication requires breaking and forming of covalent bonds. To initiate the breaking and forming of these bonds, one needs to increase the interaction strength between H (from O3$'$H) and O3$_{\alpha}$, and also between O3$'$ and P$_{\alpha}$, which have been achieved by the strategic lock-in configuration between the atoms stated above. Now, the reaction energy is the total energy needed to complete the reaction, in which the reaction will last for as long as the interaction between the reactant species are strong enough to initiate the chemical reaction. This is why the reaction energy and the activation energy do not correlate, and any correlation between them (activation and reaction energies) is just a ``lucky coincidence''. 

In any case, we have given unambiguous analyses as to why and how the A1 and A3$_{\alpha}$ chemical steps have the highest specificity and fidelity, and are precisely interdependent (with lock-in steps), which can be made compatible with contributions from Mg$^{2+}$ ions, pol-$\beta$ residues (D190 and probably others), and water molecules in the pre- and post-chemistry avenues. Our approach guarantees that the lock-in steps are only possible if H$'$ stays bonded to O3$'$ during the nucleophilic attack. 

\section{Conclusions}

In conclusion, we have actually attempted to tell an unambiguous and a self-consistent tale of how the electronic structure (with distinct energy-level spacings) may influence aging in humans using abstract analyses. These influences are found to be indirect where the unique ionization energy values determine why and how a selected set of complex biochemical reactions are initiated and executed with negligible errors to maintain healthy eukaryotes. Here, we have considered only one type of enzyme known as the polymerase-$\beta$, and a specific biochemical process involving the insertion of a single C base (from dCTP) into the growing DNA strand. Nonetheless, the above influence (due to ionization energy) is proven to be significant and far-reaching, such that the stronger renormalized vdW interaction and the Hermansson blue-shifting hydrogen bond effect are applicable to all biochemical processes. For example, these two interactions are responsible to set up the lock-in configuration that has the strongest Coulomb attraction.  

We have proven that there is only one unique biochemical process, consisting of two steps with the highest specificity and fidelity. These steps are strongly interdependent to the extent that they can be regarded as lock-in steps, occurring simultaneously. In particular, the nucleophilic attack starts when the O3$'$ uses its polarizable electron to attract the positively charged P$_{\alpha}$ by means of a stronger vdW interaction. This step leads to a stronger O3$'$---H$'$ bond, which becomes easily susceptible in the presence of O3$_{\alpha}$. At the same time, the counterattack between H$'$ and O3$_{\alpha}$ fortifies the P$_{\alpha}$---O3$_{\alpha}$ bond, which can be easily weakened by subjecting P$_{\alpha}$ to O3$'$, and this is nothing but the nucleophilic attack that is being unleashed as we speak. Obviously, these (nucleophilic- and counter-) attacks complement each other rather nicely in such a way that they lead to the formation of new O3$'$---P$_{\alpha}$ and H$'$---O3$_{\alpha}$ bonds, and the breaking of P$_{\alpha}$---O3$_{\alpha}$ and O3$'$---H$'$ bonds. In addition to this almost perfect matching, note that the bonds that are being formed are of the same types, as the ones that are being broken. This implies a negligible penalty in activation energy, or may not even require any external energy supply at all, to overcome the activation energy.

\section*{Acknowledgments}

This work was supported by Sebastiammal Innasimuthu, Arulsamy Innasimuthu, Arokia Das Anthony, Amelia Das Anthony, Malcolm Anandraj and Kingston Kisshenraj. Special thanks to Alexander Jeffrey Hinde (The University of Sydney), Mir Massoud Aghili Yajadda (CSIRO, Lindfield), Ond$\check{\rm r}$ej Grulich (Tomas Bata University, Zlin) and Naresh Kumar Mani (ENS, Paris) for providing some of the listed references.

\newpage
%%%%%%%%%%%%%%%%%%%%%%%%%%%%%%%%%%%%%%%%%%%%%%%%%%%%%%%%%%%%%%%%%%%%%%%%%%%%%%%%%%%%%%%%%%%%%%%%%%%%%%%%%%%%%%%%%%%%%
\begin{table}[ht]
%\caption{}
\begin{tabular}{l c c c} % centered columns (4 columns)
\hline\hline %inserts double horizontal lines
\multicolumn{1}{l}{Elements}        &    Atomic number &  Valence state~/    & $\xi$   \\  
\multicolumn{1}{l}{}                &   $Z$             &  $\xi$ number      & (kJmol$^{-1}$)\\  
\hline % inserts single horizontal line 

H                                   &  1   					    &  1+      & 1312 \\ 
carbon                              &  6	   	  			  &  4+      & 3571 \\ 
carbon															&  6	   	  			  &  5+      & 10423 \\
N                                   &  7                &  1+      & 1402 \\
N                                   &  7                &  2$^{\rm nd}$      & 2856 \\
N                                   &  7                &  3$^{\rm rd}$      & 4578 \\
N                                   &  7                &  4$^{\rm th}$      & 7475 \\
N                                   &  7                &  2+      & 2129 \\
N                                   &  7                &  3+      & 2946 \\
N                                   &  7                &  4+      & 4078 \\
O                                   &  8	   	  			  &  1+      & 1314 \\ 
O                                   &  8	   	  			  &  2$^{\rm nd}$      & 3388 \\ 
O                                   &  8	   	  			  &  3$^{\rm rd}$      & 5301 \\
O                                   &  8	   	  			  &  4$^{\rm th}$      & 7469 \\
O                                   &  8	   	  			  &  5$^{\rm th}$      & 10990 \\
O                                   &  8	   	  			  &  2+      & 2351 \\
O                                   &  8	   	  			  &  4+      & 4368 \\ 
O                                   &  8	   	  			  &  5+      & 5692\\ 
P                                   &  15	   	  			  &  1+      & 1012 \\ 
P                                   &  15	   	  			  &  2$^{\rm nd}$      & 1907 \\ 
P                                   &  15	   	  			  &  3$^{\rm rd}$      & 2914 \\
P                                   &  15	   	  			  &  4$^{\rm th}$      & 4964 \\
P                                   &  15	   	  			  &  5$^{\rm th}$      & 6274 \\
P                                   &  15	   	  			  &  6$^{\rm th}$      & 21267 \\
P																		&  15	   	  			  &  5+      & 3414 \\
\hline  
\end{tabular}
\label{Table:I} 
\end{table}
%%%%%%%%%%%%%%%%%%%%%%%%%%%%%%%%%%%%%%%%%%%%%%%%%%%%%%%%%%%%%%%%%%%%%%%%%%%%%%%%%%%%%%%%%%%%%%%%%%%%%%%%%%%%%%%%%%%%%%%%%%%%%%%%%%%

\textsc{Table} 1: Averaged atomic ionization energies ($\xi$) for neutral and charged elements and their respective averaged valence states ordered with increasing atomic number $Z$. All the experimental ionization energy values were obtained from Ref.~\cite{web}. The notation 2$^{\rm nd}$ denotes the 2$^{\rm nd}$ ionization energy (without averaging), and so forth. However, the valence 1+ is also equivalent to the 1$^{\rm st}$ ionization energy, while the valence state larger than 1+ has been averaged using Eq.~(\ref{eq:24b})

\begin{figure}
\begin{center}
\scalebox{0.5}{\includegraphics{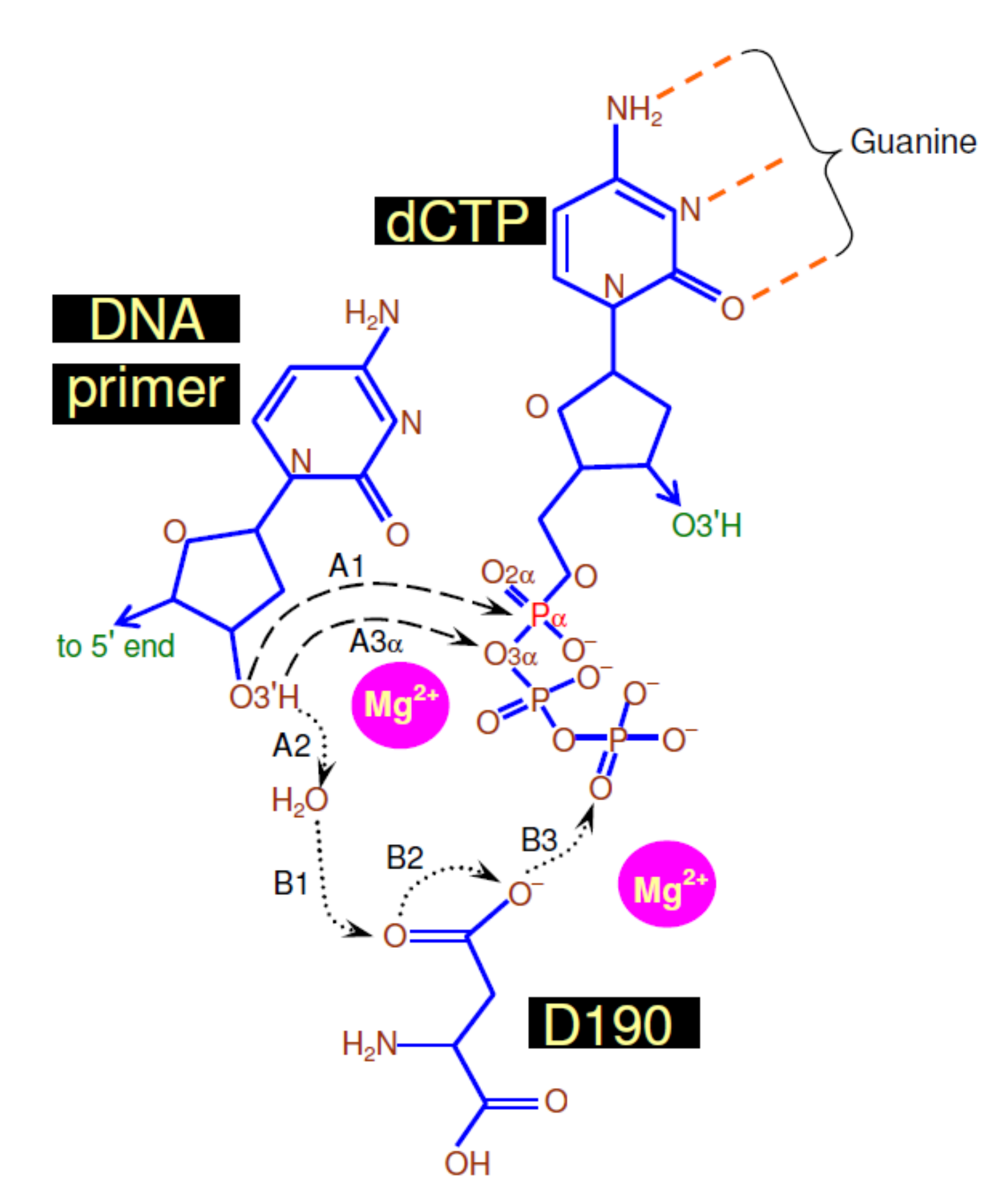}}
\caption{}
\label{fig:2}
\end{center}
\end{figure}

\newpage
%%%%%%%%%%%%%%%%%%%%%%%%%%%%%%%%%%%%%%%%%%%%%%%%%%%%%%%%%%%%%%%%%%%%%%%%%%%%%%%%%%%%%%%%%%%%%%%%%%%%%%%%%%%%%%%%%%%%%%%%%%%%%
\textsc{Figure} 1: Two amino acids (the carboxylic-group from the amino acid 1 and the amine-group from the amino acid 2) react to form a dipeptide and a water molecule. We have identified three different hydrogen atoms, H$_{\rm I}$, H$_{\rm II}$ and H$_{\rm III}$, two different carbon atoms (C$_{\rm I}$ and C$_{\rm II}$) and two different oxygen atoms, namely, O$_{\rm I}$ and O$_{\rm II}$. The organic substituent, $\texttt{R}$ is irrelevant in this reaction for obvious reasons. Dashed arrows indicate the correct reaction pathways (NH$_{\rm I}$ $\rightarrow$ CO$_{\rm I}$), while the alternative reaction pathways are between these pairs of atoms, O$_{\rm I}$H$_{\rm III}$(from amino acid 2) $\rightarrow$ O$_{\rm I}$H$_{\rm III}$(from amino acid 1), NH$_{\rm I}$(from amino acid 2) $\rightarrow$ O$_{\rm I}$H$_{\rm III}$(from amino acid 1), O$_{\rm I}$H$_{\rm III}$(from amino acid 2) $\rightarrow$ CO$_{\rm I}$(from amino acid 1) and CH$_{\rm II}$(from amino acid 2) $\rightarrow$ CO$_{\rm I}$(from amino acid 1). The above notations imply that the first pair (of atoms) belongs to the acid amino 2, whereas the second pair (on the right-hand side of ``$\rightarrow$'') is from the acid amino 1. These alternatives will not be discussed---see text for details.
%%%%%%%%%%%%%%%%%%%%%%%%%%%%%%%%%%%%%%%%%%%%%%%%%%%%%%%%%%%%%%%%%%%%%%%%%%%%%%%%%%%%%%%%%%%%%%%%%%%%%%%%%%%%%%%%%%%%%%%%%%%%
\\ \\
%%%%%%%%%%%%%%%%%%%%%%%%%%%%%%%%%%%%%%%%%%%%%%%%%%%%%%%%%%%%%%%%%%%%%%%%%%%%%%%%%%%%%%%%%%%%%%%%%%%%%%%%%%%%%%%%%%%%%%%%%%%%%
\textsc{Figure} 2: There are two processes involved for a single base (C) insertion into the 3$'$ end of the DNA primer, depicted in this schematic figure. The first process has been proposed in Refs.~\cite{rad,alb} that consists of five steps, A1 to B3. Whereas, the second process proven here involves A1 and A3$_{\alpha}$ occurring simultaneously (follow the dashed arrows). The first two major steps (discovered by Radhakrishnan and Schlick~\cite{rad}) carry the labels, A1 and A2, while the other steps (from B1 to B3) are part of the subsequent proton(H$^+$)-transfer steps assisted by D190, proposed by Alberts, Wang and Schlick~\cite{alb} (follow the dotted arrows). The diagonal dashed lines on the top right corner denote the possible Watson-Crick hydrogen bonds with a guanine (from a template strand), while the solid diagonal arrow pointing to O3$'$H in an incoming nucleotide (dCTP) will be the new 3$'$ end of the primer (growing) strand, after insertion. Another diagonal arrow pointing to the left (in the DNA primer) extends in the 5$'$ end direction. All the sticks (including the isolated sticks inside the ring) and the diagonal solid arrows represent the covalent $\sigma$ bonds. The conformations and the positions of the DNA primer, D190, dCTP and Mg$^{2+}$ are not to scale. The number 190 refers to the position of an Aspartate residue attached to a pol-$\beta$ enzyme.
%%%%%%%%%%%%%%%%%%%%%%%%%%%%%%%%%%%%%%%%%%%%%%%%%%%%%%%%%%%%%%%%%%%%%%%%%%%%%%%%%%%%%%%%%%%%%%%%%%%%%%%%%%%%%%%%%%%%%%%%%%%%
\\ \\
%%%%%%%%%%%%%%%%%%%%%%%%%%%%%%%%%%%%%%%%%%%%%%%%%%%%%%%%%%%%%%%%%%%%%%%%%%%%%%%%%%%%%%%%%%%%%%%%%%%%%%%%%%%%%%%%%%%%%%%%%%%%%%%
\textsc{Figure} 3: The above schematic diagrams are sketched to depict the nucleophilic-attack region, selected and magnified from Fig.~\ref{fig:2} for different types of counterattack (counter to the nucleophilic attack) between oxygen (bonded to P$_{\alpha}$) and H$'$. The dotted arrows indicate the induced electrostatic interactions between $\delta +$ and $\delta -$ charges for different oxygen atoms bonded to P$^{\delta +}_{\alpha}$, for instance, (A) $^{\delta -}$O3$_{\alpha}$ (also bonded to another P), (B) $^{\delta -}$O2$_{\alpha}$, (C) O$^{\delta -}$ (also bonded to carbon) and (D) O$^-$. Note here that the nucleophilic attack (between $^{\delta -}$O3$'$ and P$^{\delta +}_{\alpha}$) remains the same. The molecular conformations and their positions are not to scale.
%%%%%%%%%%%%%%%%%%%%%%%%%%%%%%%%%%%%%%%%%%%%%%%%%%%%%%%%%%%%%%%%%%%%%%%%%%%%%%%%%%%%%%%%%%%%%%%%%%%%%%%%%%%%%%%%%%%%%%%%%%%%

\newpage
%%%%%%%%%%%%%%%%%%%%%%%%%%%%%%%%%%%%%%%%%%%%%%%%%%%%%%%%%%%%%%%%%%%%%%%%%%%%%%%%%%%%%%%%%%%%%%%%%%%%%%%%%%%%%%%%%%%%%%%%%%%%%
%%%%%%%%%%%%%%%%%%%%%%%%%%%%%%%%%%%%%%%%%%%%%%%%%%%%%%%%%%%%%%%%%%%%%%%%%%%%%%%%%%%%%%%%%%%%%%%%%%%%%%%%%%%%%%%%%%%%%%%%%%%%

\newpage
%%%%%%%%%%%%%%%%%%%%%%%%%%%%%%%%%%%%%%%%%%%%%%%%%%%%%%%%%%%%%%%%%%%%%%%%%%%%%%%%%%%%%%%%%%%%%%%%%%%%%%%%%%%%%%%%%%%%%%%%%%%%%
%%%%%%%%%%%%%%%%%%%%%%%%%%%%%%%%%%%%%%%%%%%%%%%%%%%%%%%%%%%%%%%%%%%%%%%%%%%%%%%%%%%%%%%%%%%%%%%%%%%%%%%%%%%%%%%%%%%%%%%%%%%%

\newpage
%%%%%%%%%%%%%%%%%%%%%%%%%%%%%%%%%%%%%%%%%%%%%%%%%%%%%%%%%%%%%%%%%%%%%%%%%%%%%%%%%%%%%%%%%%%%%%%%%%%%%%%%%%%%%%%%%%%%%%%%%%%
%%%%%%%%%%%%%%%%%%%%%%%%%%%%%%%%%%%%%%%%%%%%%%%%%%%%%%%%%%%%%%%%%%%%%%%%%%%%%%%%%%%%%%%%%%%%%%%%%%%%%%%%%%%%%%%%%%%%%%%%%%%%

\end{document}